\documentclass[pra,twocolumn,floatfix,superscriptaddress]{revtex4}
\usepackage{pstricks,graphicx,amsmath,bbm,mathrsfs,amssymb,psfrag,pifont,times,mathptmx}
\usepackage[utf8]{inputenc}
\usepackage{subfigure}
\usepackage[english]{babel}
\usepackage{color}

\DeclareMathOperator{\Tr}{Tr}
\begin{document}
\title{Squeezing-enhanced phase-shift-keyed binary communication in noisy channels}
\author{Giovanni Chesi}
\affiliation{Dipartimento di Fisica e Matematica, Universit\`a degli Studi dell'Insubria, I-22100 Como, Italy}
\author{Stefano Olivares}
\email{stefano.olivares@fisica.unimi.it}
\affiliation{Quantum Technology Lab, Dipartimento di Fisica ``Aldo Pontremoli'', Universit\`a degli Studi di Milano, I-20133 Milano, Italy}
\affiliation{INFN, Sezione di Milano, I-20133 Milano, Italy}
\author{Matteo G. A. Paris}
\affiliation{Quantum Technology Lab, Dipartimento di Fisica ``Aldo Pontremoli'', Universit\`a degli Studi di Milano, I-20133 Milano, Italy}
\affiliation{INFN, Sezione di Milano, I-20133 Milano, Italy}
\date{\today}
%%%%%%%%%%%%%%
\begin{abstract}
We address binary phase-shift-keyed communication channels based 
on  Gaussian states and prove that squeezing improves state discrimination 
at fixed energy of the channel, also in the presence of phase diffusion.
We then assess performances of homodyne detection against the ultimate 
quantum limits to discrimination, and show that homodyning achieves optimality 
in large noise regime. Finally, we consider noise  in the preparation of the 
{\em seed} signal (before phase encoding) and show that also in this
case squeezing may improve state discrimination in realistic conditions.
\end{abstract}
\maketitle
%%%
\section{Introduction}
Quantum hypothesis testing addresses the discrimination of non-orthogonal 
preparations of a quantum system. As such, it has attracted attention 
from a fundamental point of view \cite{YKL,helstrom,r1,r2,r3,IDP1,IDP2,IDP3,tom}, 
and represents a central tool for the design and development of quantum 
communication channels.
In particular, much attention has been payed to quantum binary discrimination,
which already shows a rich quantum phenomenology amenable to analytic 
investigations, and fosters promising perspectives in a number of 
challenging applications, such as quantum communication \cite{cariolaro:15,leuchs:17}
and quantum cryptography \cite{diamanti:npj:16}.
In particular, quantum-optical implementations of state discrimination
have been investigated \cite{becerra1,wittmann,muller}, with focus on
information carried by coherent states \cite{becerra2, nair}, though also 
the use of squeezing have been explored to some extent \cite{p01,izumi}.
\par
In quantum binary communication, one has to discriminate between
two quantum signals. Assuming a pure preparation, i.e. ignoring the 
presence of noise, we may denote the two states by $|\psi_1\rangle$ and 
$|\psi_2\rangle$. The minimum error probability is not vanishing 
for non-orthogonal states and is given by the so-called Helstrom 
bound \cite{helstrom} (throughout the paper we assume equal prior probability
for the two signals):
\begin{equation}\label{hel:bound}
P^{\rm (min)} (|\psi_1\rangle, |\psi_2\rangle) = \frac{1}{2}(1-\sqrt{1-|\langle\psi_1 | \psi_2\rangle|^2}).
\end{equation}
In quantum communication channels the two states are often obtained 
from a {\it seed} state $| \psi_0 \rangle$ through a suitable unitary transformation.
In coherent-state communication, the quantum signal 
$| \psi_k \rangle$, $k=1,2$, is obtained by applying a
displacement operation to the vacuum state $| 0 \rangle$, namely,
\begin{equation}
| \psi_k \rangle =  {D}\left[ (-1)^k \alpha \right] | 0 \rangle\,,\quad (k=1,2)
\end{equation}
where $ {D}(\alpha)=\exp(\alpha  {a}^{\dagger}-\alpha^{*} {a})$,
$[ {a}, {a}^{\dagger}] =  {\mathbbm I}$
and we can assume $\alpha \in {\mathbbm R}$, $\alpha >0$. Since 
the difference between the two states is an overall $\pi$ phase, we 
refer to this encoding as to phase-shift keying (PSK).
This channel has been thoroughly investigated because of the
peculiar properties of coherent states: they can be easily generated and manipulated and 
their phase properties are not affected by losses (only their amplitude is reduced). Therefore, they
can travel for long distances preserving their coherence 
properties \cite{leuchs:17,lau:06}.
\par
The search for optimal receivers, i.e. detection schemes and 
strategies able to discriminate between the two coherent states 
reaching the corresponding Helstrom bound, has eventually led to
the so-called {\itshape Kennedy receiver} \cite{kennedy} and 
{\itshape Dolinar receiver} \cite{dolinar}. These are both based 
on the interference of the signals with a known reference, and on 
on/off photo-detectors, able to check the presence or absence of 
light. Dolinar receiver also exploits a feedback mechanism
in order to achieve the Helstrom bound (it requires, however, many
copies of the same input state in order to implement the feedback).
These kinds of detectors require a precise phase reference control and in the
presence of phase fluctuations their performances are drastically reduced \cite{bina:16}.
Nevertheless, it has been shown that in the presence of phase noise
a receiver based on homodyne detection \cite{kazovsky} allows to approach optimality \cite{olivares2}.
\par
Recently, coherent-state encoding 
and homodyne detection have been used to establish ground-satellite 
links \cite{leuchs:17}, whereas squeezed states have been suggested 
to improve quantum key distribution \cite{vlad:11}. Motivated by these results, 
here we consider a binary communication channel in which the seed
$| \psi_0 \rangle$ is a squeezed vacuum $| r \rangle =  {S}(r)| 
0 \rangle$ where $ {S}(r)=\exp\{\frac{1}{2}[r ( {a}^
{\dagger})^2 - r^{*} {a}^2]\}$ is the squeezing operator.
Without lack of generality we assume $r \in {\mathbbm R}$. In 
this case the input states are:
\begin{equation}\label{DSS:inputs}
| \psi_1 \rangle =  | -\alpha, r \rangle
\quad \mbox{and} \quad
| \psi_2 \rangle = | \alpha, r \rangle,
\end{equation}
where we introduced the displaced squeezed state (DSS)
\begin{equation}
| \alpha, r \rangle =  {D}(\alpha)  {S}(r) |  0 \rangle\,.
\end{equation}
In the following, we will first address the Helstrom bound for two DSSs and compare
the results with those obtained with coherent states and/or 
using homodyne detection. We then investigate
the effect of phase diffusion on the discrimination and 
compare our results with the ultimate quantum limits, i.e. 
the corresponding Helstrom bound. Finally, we analyze the effect of 
losses, resulting in a reduced purity of the seed state.
\section{Discrimination between displaced-squeezed states}\label{sec:basics}
Let us start by investigating the ultimate performances of DSSs with 
respect to the coherent-state ones in the absence of noise and with optimal
detection. This corresponds to evaluate the Helstrom bound for
the input states given in Eqs.~(\ref{DSS:inputs}).
For the sake of simplicity,
we can assume $\alpha,r \in {\mathbbm R}$, with $\alpha,r >0$. The corresponding Helstrom bound can be easily calculated
from Eq.~(\ref{hel:bound}) and reads:
\begin{align}
P^{({\rm min})} (r,\alpha) & = \frac{1}{2}\left(1-\sqrt{1-|\langle\alpha,r|-\alpha,r\rangle|^2}\right) \nonumber \\
& =  \frac{1}{2}\left[1-\sqrt{1-\exp(-4\alpha^2e^{-2r})}\right]. \label{HB:DSS}
\end{align}
It is clear that for fixed coherent amplitude, squeezing always allows to improve the discrimination. However,
the squeezing operation is adding energy to the coherent states, resulting in a comparison between
two encodings with different total energy. A better comparison may be obtained by fixing the {\rm total energy}
used in the communication channel which is usually set by physical constraints.
For the coherent-state channel based on $| \pm\alpha \rangle$ we have the
following average number of photons $N_{\rm CS} = |\pm\alpha|^2$, whereas, if we employ DSSs $| \pm \alpha, r \rangle$,
the energy is given by $N_{\rm DSS} = |\alpha|^2+ N_{\rm sq}$, where $N_{\rm sq} = \sinh^2(r)$ is the average number of
photons added by the squeezing process.
If we impose the constraint on the energy, we can clearly see that the more the state is squeezed the less is displaced and vice versa;
hence, the discrimination problem becomes not trivial.
\par
\begin{figure}[h!]
\includegraphics[width=0.495\columnwidth]{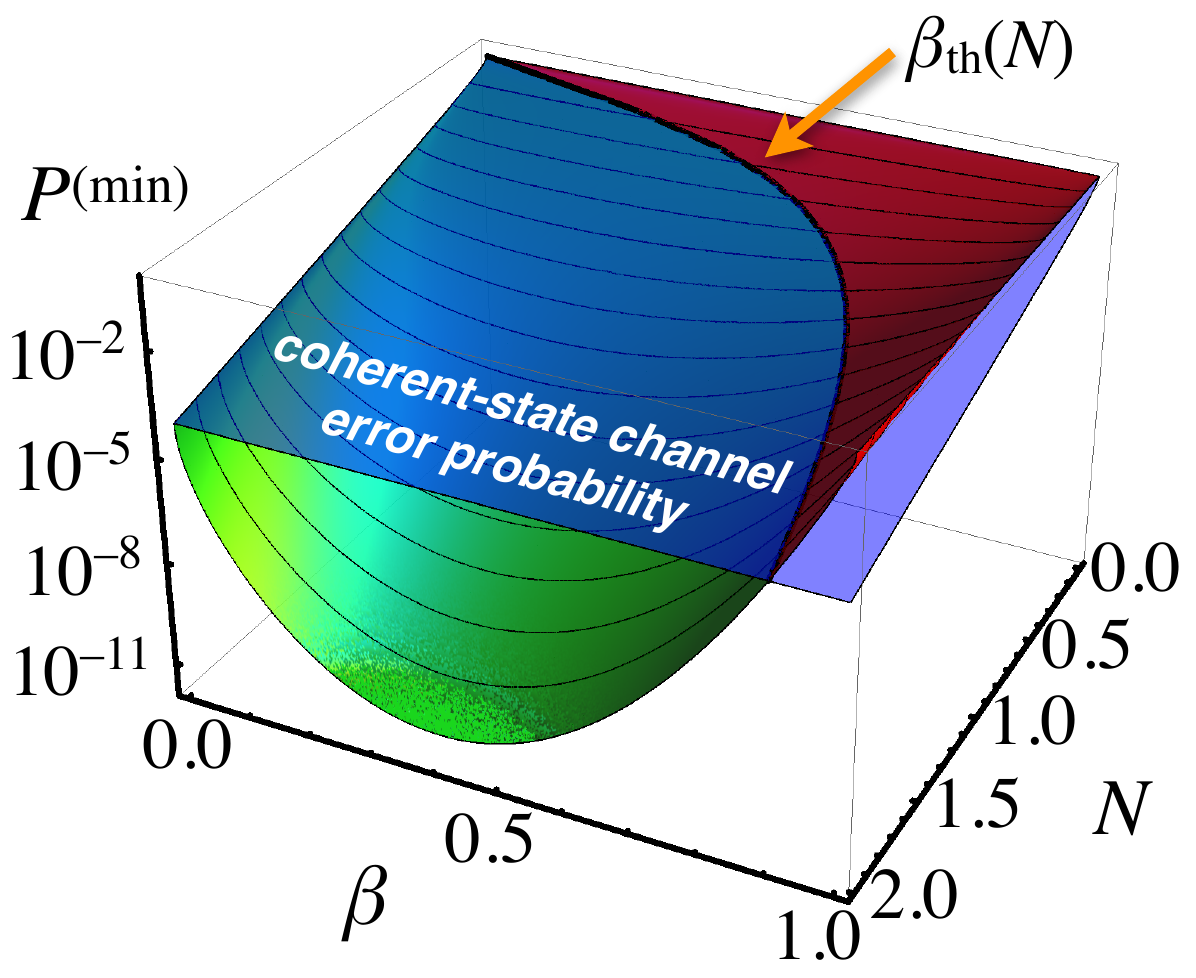}
\includegraphics[width=0.495\columnwidth]{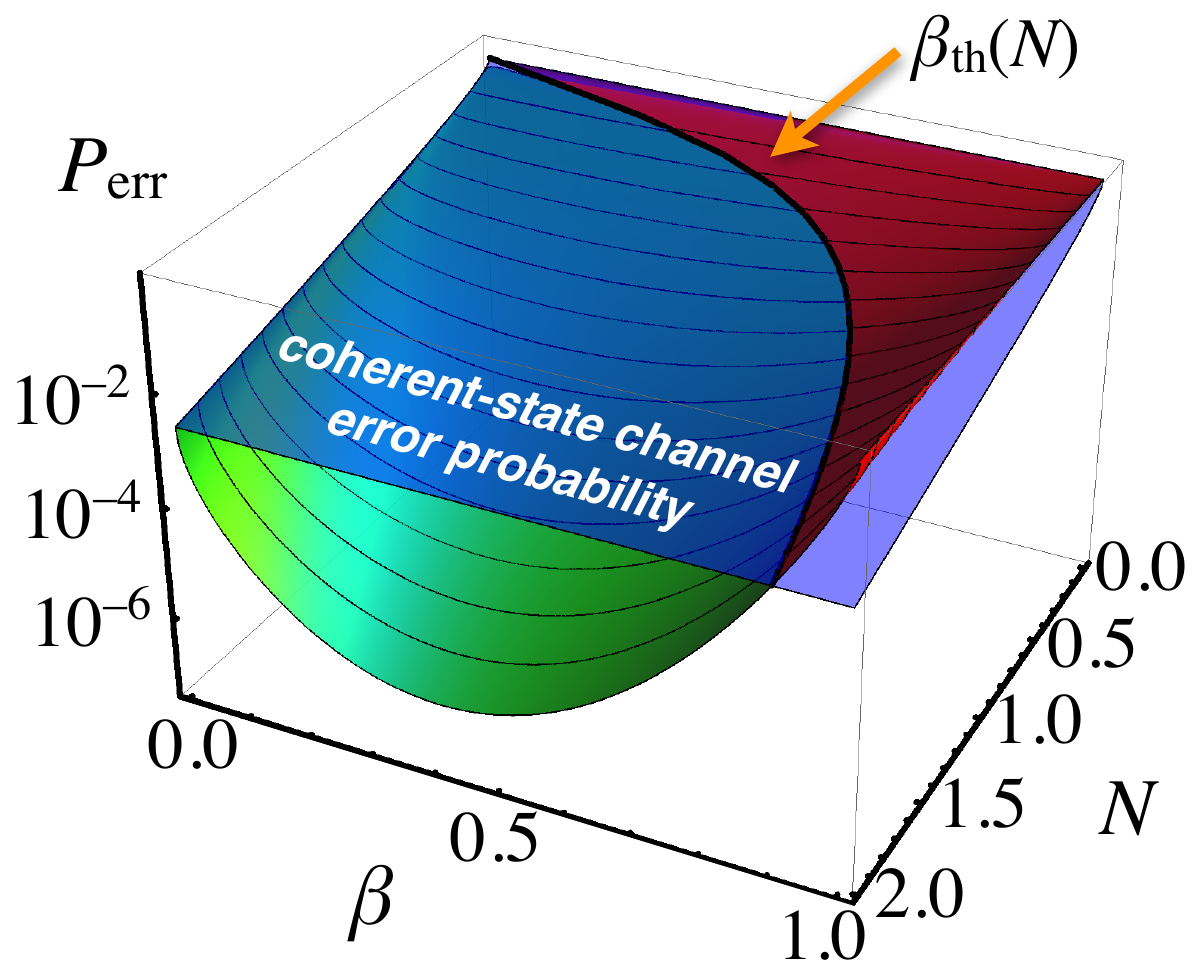}
\caption{(Left) The Helstrom bound (\ref{HB:param})
as functions of the squeezing fraction $\beta$ and the channel 
energy $N$. We also plot the plane corresponding to the Helstrom bound 
for coherent states. The solid line corresponds to the threshold 
$\beta_{\rm th}(N)$: given the total energy $N$, if $\beta < 
\beta_{\rm th}(N)$ the use of DSSs outperforms coherent states.
(Right) The homodyne error probability $P_{\rm err}$ of 
Eq.~(\ref{Perr:id:HD}) as a function of $\beta$
for different values of the channel energy $N$. The behavior is qualitatively similar to the Helstrom bound reported in the left panel (see however 
the different scaling). We also plot the plane corresponding 
to the minimum error probability
achievable using only coherent states and homodyne detection.}
\label{f:hbper}
\end{figure}

Our aim is to find the regimes in which squeezing can be a useful resource for the discrimination.
Therefore, in the following we are going to study the error probability as a function of the introduced amount of squeezing
(and at fixed total energy).  Given the channel energy $N$ and the fraction of squeezing:
\begin{equation} \label{sq:frac}
\beta \equiv \frac{N_{\rm sq}}{N}=\frac{\sinh^2(r)}{N},
\end{equation}
we can rewrite Eq.~(\ref{HB:DSS}) expressed in terms of $\beta$ and $N$, namely:
\begin{align}
P^{\rm (min)}(\beta, N)
=\frac{1}{2}&\, - \frac12\bigg\{1-\exp\Big[-4N(1-\beta)  \nonumber \\
& \times\Big(1 + 2N\beta+2\sqrt{N\beta(1+N\beta)}\Big)\Big]  \bigg\}^{\frac12}.
\label{HB:param}
\end{align}
We plot the behavior of $P^{\rm (min)}(\beta, N)$ in the left panel of 
Fig.~\ref{f:hbper}. We can find an interval of $\beta$ values
such that the error probability of a pair of DSSs is smaller than the corresponding coherent case. Then two critical values for $\beta$ are naturally identified: a threshold value $\beta_{\rm th}(N)$ and an optimum value $\beta_{\rm opt}$. For $0<\beta<\beta_{\rm th}$, DSSs achieve better performances with respect to coherent states and, in particular, it is possible to minimize the error probability by choosing a suitable squeezing fraction ($\beta_{\rm opt}$). These quantities can be obtained analytically and reads:
\begin{equation}\label{beta:th:opt}
\beta_{\rm th}(N)  =  \frac{4N}{4N+1}, \quad \text{and} \quad
\beta_{\rm opt}(N)  =  \frac{N}{2N+1},
\end{equation}
respectively. It is straightforward to see that in the limit $N \gg 1$ one finds $\beta_{\rm th}\rightarrow 1$ and $\beta_{\rm opt} \rightarrow 1/2$.
In this limit, the best strategy is to use the half of the channel energy for squeezing. Moreover, we have that, for coherent states ($\beta=0$), $P^{\rm (min)}_{\rm CS} = P^{\rm (min)}(0,N) = \frac{1}{2}\left[1-\sqrt{1-\exp(-4N)}\right]$, whereas for DSSs with $\beta = \beta_{\rm opt}$ we find $P^{\rm (min)}_{\rm DSS} = P^{\rm (min)} (\beta_{\rm opt},N) = \frac{1}{2}\left[1-\sqrt{1-\exp[-4N(N+1)]}\right]$.
Hence, in the limit $N \gg 1$ the corresponding minimum error probabilities are:
\begin{subequations}
\label{limit:P:min}
\begin{align}
P^{\rm (min)}_{\rm CS} & \rightarrow \frac{1}{4}\exp(-4N), \\
P^{\rm (min)}_{\rm DSS} & \rightarrow \frac{1}{4}\exp[-4N(N+1)] \sim \frac{1}{4}\exp(-4N^2),
\end{align}
\end{subequations}
and the clear advantage can be also highlighted by considering the following ratio:
\begin{equation}
\frac{P^{\rm (min)}_{\rm CS}(0,N)  -P^{\rm (min)}_{\rm DSS}(\beta_{\rm opt},N) }
{P^{\rm (min)}_{\rm CS}(0,N) }\rightarrow 1-\exp(-4N^2).
\label{sqsupremacy}
\end{equation}
 \\

Let us now turn  attention on feasible measurements that we can perform
on our optical states in order to discriminate between them. It is 
well-known that in the absence of noise and in the case of just 
two phase-shifted coherent states there exist methods based on 
photon-number resolving detectors which allow one to approach 
the Helstrom bound \cite{kennedy,dolinar,izumi12,izumi16}.
Other proposals are based on an active detection stage, namely, 
a squeezing operation is applied to the signal before the 
detection \cite{izumi}.
\par
However, one has to take into account noise to describe a 
realistic channel. Since we chose to employ a PSK channel, 
information is encoded on phase, so that the most detrimental
kind of noise affecting our channel is phase noise. It has 
been shown \cite{olivares2} that homodyne detection is robust 
against phase noise; in particular, in the limit of large 
noise it beats  the performances of a Kennedy receiver for every value 
of $N$ and achieves the Helstrom bound. These results lead us to
consider homodyne detection in order to discriminate DSSs.
On the one hand, this is a standard and well-established technique 
when dealing with continuous-variable optical
system, which is also promising in view of coherent-state
communication with satellite \cite{leuchs:17}.
On the other hand, homodyne detection strategy has been 
proved to approach the Helstrom bound when phase
noise affect the signal \cite{olivares2,bina17}, which 
is the scenario we will discuss in the next sections.
\par
Homodyne detection allows to measure the quadrature operator
$ {x}_{\theta}\equiv {a}e^{-i\theta}+ {a}^{\dagger}e^{i\theta}$
of the input field \cite{bachor}, where $[ {a}, {a}^{\dag}] = {\mathbbm I}$ and $\theta$
is the phase of the quadrature. Since we assumed real displacement amplitude and squeezing,
we can focus on the measurement of $ {x}_{0}$.
As usual the two inputs $| \pm \alpha, r \rangle$, with $\alpha, r > 0$, are sent to the receiver,
which measures $ {x}$ and obtain the outcome $x \in {\mathbbm R}$ from the homodyne detection.
In order to discriminate between the two states, shot by shot, the receiver uses the following
strategy:
\begin{equation}\left\{
\begin{array}{lcl}
x\ge 0 & \Rightarrow & | 0 \rangle \equiv | \alpha, r \rangle, \\
x< 0 & \Rightarrow &  | 1 \rangle \equiv | -\alpha, r \rangle.
\end{array}
\right.
\end{equation}
The error probability associated with this strategy writes:
\begin{equation}\label{Perr:id:HD}
P_{\rm err}=\frac{1}{2}\big[p(x \ge 0 | 1)+p(x < 0 | 0)\big] = p(x \ge 0 | 1).
\end{equation}
where we used $p(x < 0 | 0) = p(x \ge 0 | 1)$.
The conditional probabilities appearing in Eq.~(\ref{Perr:id:HD}) are defined as:
\begin{equation}
p(x \ge 0 | 1) = \int_{0}^{+\infty} dx\, p^{\rm (HD)}(x; \pm \alpha, r ),
\end{equation}
where $p^{\rm (HD)}(x; \pm \alpha, r )$ is the {\em homodyne probability}, which reads:
\begin{align}
p^{\rm (HD)}(x; \pm \alpha, r ) & \equiv | {}_0\langle x | \pm \alpha, r \rangle |^2 \nonumber \\
&= \frac{1}{\sqrt{2\pi\, e^{-2r}}}\, \exp \left\{-\frac{[(x \pm 2\alpha)]^2}{2\, e^{-2r}}\right\},
\end{align}
with $ {x}_\theta | x \rangle_\theta = x | x \rangle_\theta$, or, in terms of the squeezing
fraction $\beta$ and the total number of photons $N$, as
\begin{equation}
p^{\rm (HD)}(x; \beta, N ) =
\frac{\exp\left\{
{\displaystyle -\frac{\Big[x\pm2\sqrt{N(1-\beta)}\Big]^2}{2\, \Sigma^2}}
\right\}}
{\sqrt{2\,\pi\,\Sigma^2}},
\label{perr2}
\end{equation}
with $\Sigma = \Sigma(\beta,N) = (\sqrt{N \beta}+\sqrt{1 + N \beta})^{-1}$.
If we consider the regime $N \gg 1$, we obtain the following expressions
of the error probabilities for coherent states ($\beta = 0$):
\begin{equation}
P_{\rm  err,\, CS} \to \frac14 \sqrt{\frac{2}{\pi}} \frac{e^{-2N}}{\sqrt{N}}\,,
\end{equation}
and for DSSs:
\begin{equation}
P_{\rm  err,\, DSS} \to \frac14 \sqrt{\frac{2}{\pi}} \frac{e^{-2N^2}}{N}\,,
\end{equation}
respectively, where, in the latter we used $\beta = \beta_{\rm opt}$ given in
Eqs.~(\ref{beta:th:opt}). It is interesting to note that we find a dependence on
$N$ for coherent states and on $N^2$ for the DDSs, as we obtained
for the Helstrom bound in Eqs.~(\ref{limit:P:min}), but, now, with a
different scaling.
\par
In the right panel of Fig.~\ref{f:hbper} we show  $P_{\rm err}$ as a 
function of the squeezing fraction $\beta$ and $N$. The plot well illustrates
two features of the detection strategy. On the one hand, PSK based on 
DSSs outperforms the corresponding coherent protocol as far as the squeezing
fraction is below a threshold value, which is the same minimising the 
Helstrom bound i.e. that given in Eqs.~(\ref{beta:th:opt}). Notice that
being the encoding on pure states, the same working regime is also maximising the mutual information between the sender and the receiver \cite{tom}.
On the other hand, upon comparing the homodyne error probability with the ultimate bound of 
the left panel, we see that homodyne detection does not implement the optimum receiver, since the error probability is far from approaching the Helstrom bound. At the same time, it is known that homodyne performances with 
coherent encoding are useful in the presence of phase noise and, in particular, phase diffusion \cite{olivares2,genoni}. Therefore, in the following, 
we are going to investigate whether the use of squeezing can further improve the discrimination between the two signals when phase noise affects their 
propagation.
\section{Discrimination in the presence of phase noise}\label{sec:noise}
The effect of phase diffusion on a single-mode state can be described by the the following Master equation \cite{genoni,gardiner}: 
\begin{equation}
\dot{ {\rho}}(t) = \Gamma {\cal L}[ {a}^{\dagger} {a}] {\rho}(t)
\label{me}
\end{equation}
where $ {\rho}(t)$ is the density operator of the system, 
${\cal L}[ {O}] {\rho} = 2 {O} {\rho} {O}^{\dagger}- {O}^{\dagger} {O} {\rho} -  {\rho} {O}^{\dagger} {O}$,
and $\Gamma$ is the diffusion parameter. Simple calculations provide the following solution
\begin{equation}
{\rho}(t) = \sum_{n,m} \rho_{n,m}(t)|n\rangle\langle m| = \sum_{n,m}e^{-(n-m)^2\Delta}\rho_{n,m}(0)|n\rangle\langle m|
\label{solution}
\end{equation}
where $ {\rho}_{n,m}(t) = \langle n |  {\rho}(t) | m\rangle$ and $\Delta \equiv \Gamma t$.
\begin{figure}[h!]
\includegraphics[width=0.51\columnwidth]{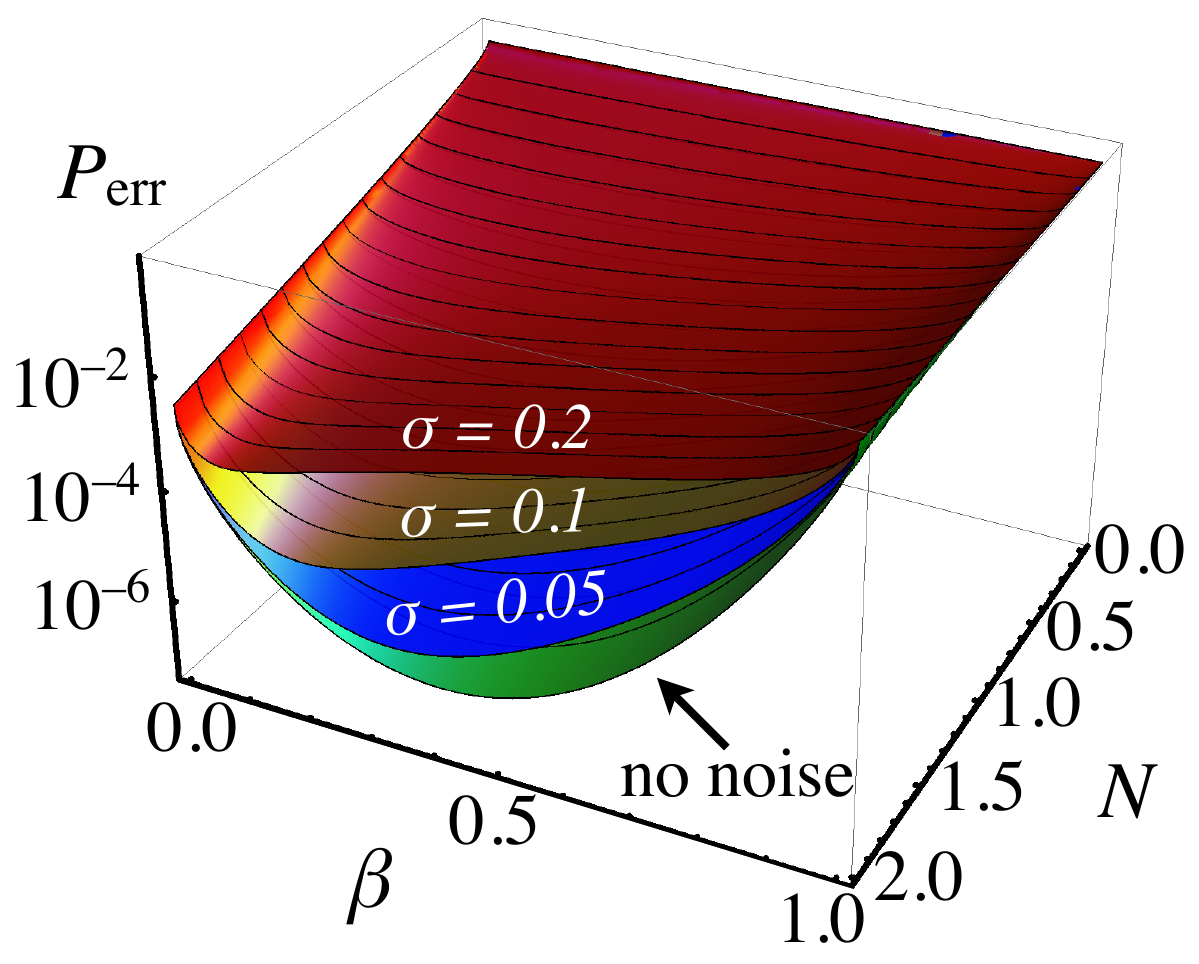}
\includegraphics[width=0.45\columnwidth]{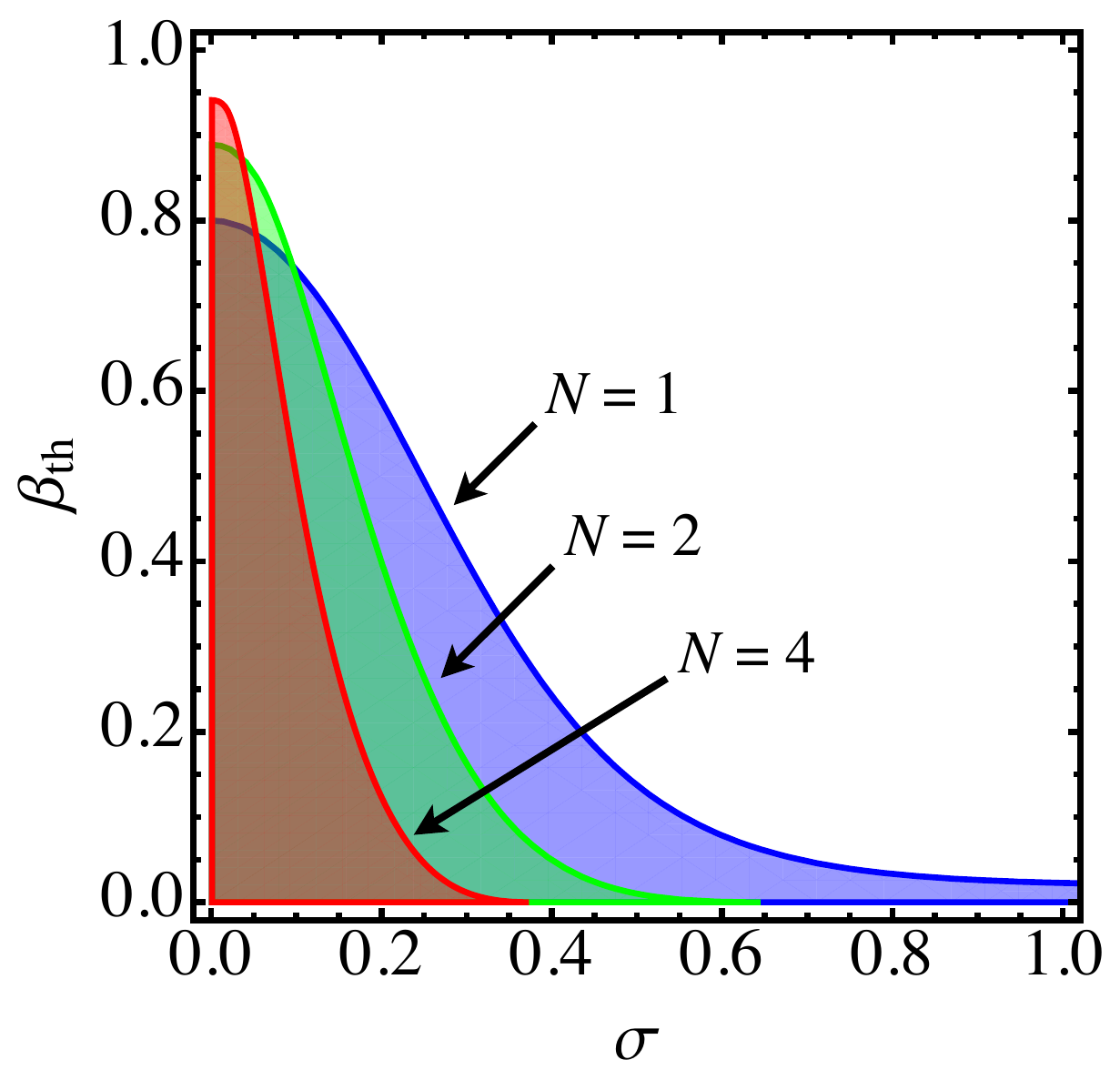}
\caption{(Left) Error probability $P_{\rm err}$ of the 
homodyne receiver in the presence of phase diffusion as 
a function of $\beta$ and $N$ for different values 
of the noise parameter $\sigma$. (Right) Threshold value $\beta_{\rm th}(\sigma)$ as a function of the noise parameter $\sigma$ for different
values of $N$. The shaded regions refer to the couples of parameters $(\sigma,\beta)$ such that DDSs outperform
coherent states.} 
\label{fig:Perr:HD:noise}
\end{figure}

Equation~(\ref{solution}) can be rewritten in the following form:
\begin{equation}
{\rho}(t) = \int_{-\infty}^{+\infty} d\phi \, \frac{\exp\left(-\frac{\phi^2}{2\sigma^2}\right)}{\sqrt{2\pi\, \sigma^2}}\,
{U}_{\phi} {\rho}(0) {U}_{\phi}^{\dagger}\,,
\label{solution2}
\end{equation}
where we introduced the phase shift operator $ {U}_{\phi}=e^{-i \phi  {a}^{\dagger} {a}}$,
$\sigma^2 = 2 \Gamma t = 2 \Delta$ and $ {\rho}(0) = |\pm\alpha,r\rangle  \langle\pm\alpha,r\rangle |$.
From now on we will refer to $\sigma$ as to the noise parameter.
\par
Since ${U}_{\phi}|\pm\alpha,r\rangle = |\pm\alpha e^{-i\phi}, r e^{-i2\phi} \rangle$,
one can easily calculate the error probability after the phase diffusion process starting from the homodyne probabilities:
\begin{align}
p^{\rm (HD)}_{\sigma}(x; \pm\alpha, r ) =
\int^{+\infty}_{-\infty} d\phi\, &\frac{\exp\left(-\frac{\phi^2}{2\sigma^2}\right)}{\sqrt{2\pi\, \sigma^2}}
\nonumber \\
&\times | {}_0\langle x | \pm \alpha e^{-i\phi}, r e^{-i\phi} \rangle |^2\,.
\end{align}
\par
In Fig.~\ref{fig:Perr:HD:noise} (left panel) we plot the resulting error probability 
$P_{\rm err}(\beta,N;\sigma)$ as a function of $\beta$ and $N$
for different values of the noise parameter $\sigma$: as one may expect, the error probability increases but, nevertheless,
the DSSs perform better than the coherent states when $\beta$ is below a new threshold $\beta_{\rm th}(N,\sigma)$ that now depends on both $N$ and the noise parameter $\sigma$, as shown in the right panel of Fig.~\ref{fig:Perr:HD:noise}.

\subsection{Comparison with the ultimate bounds}\label{sec:ultimate:noise}
Here we compare the results obtained in the previous section with the minimum error probability given by the Helstrom bound
in the presence of phase noise. Given two mixed states $ {\rho}_{1}$ and $ {\rho}_{2}$, the Helstrom bounds reads \cite{helstrom}:
\begin{equation}
P^{\rm (min)}( {\rho}_1, {\rho}_2)=\frac{1}{2}\big[1-D( {\rho}_{1}, {\rho}_{2})\big],
\label{eptd}
\end{equation}
where $D( {\rho}_{1}, {\rho}_{2})\equiv\frac{1}{2}\| {\rho}_{1}- {\rho}_{2}\|_{1}$, with
$\| {A}\|_{1}=\frac{1}{2}\Tr\left[\sqrt{ {A}^{\dagger} {A}}\right]$,
is the trace distance between the states $ {\rho}_{1}$ and $ {\rho}_{2}$.
\par
\begin{figure}[h!]
\includegraphics[width=0.95\columnwidth]{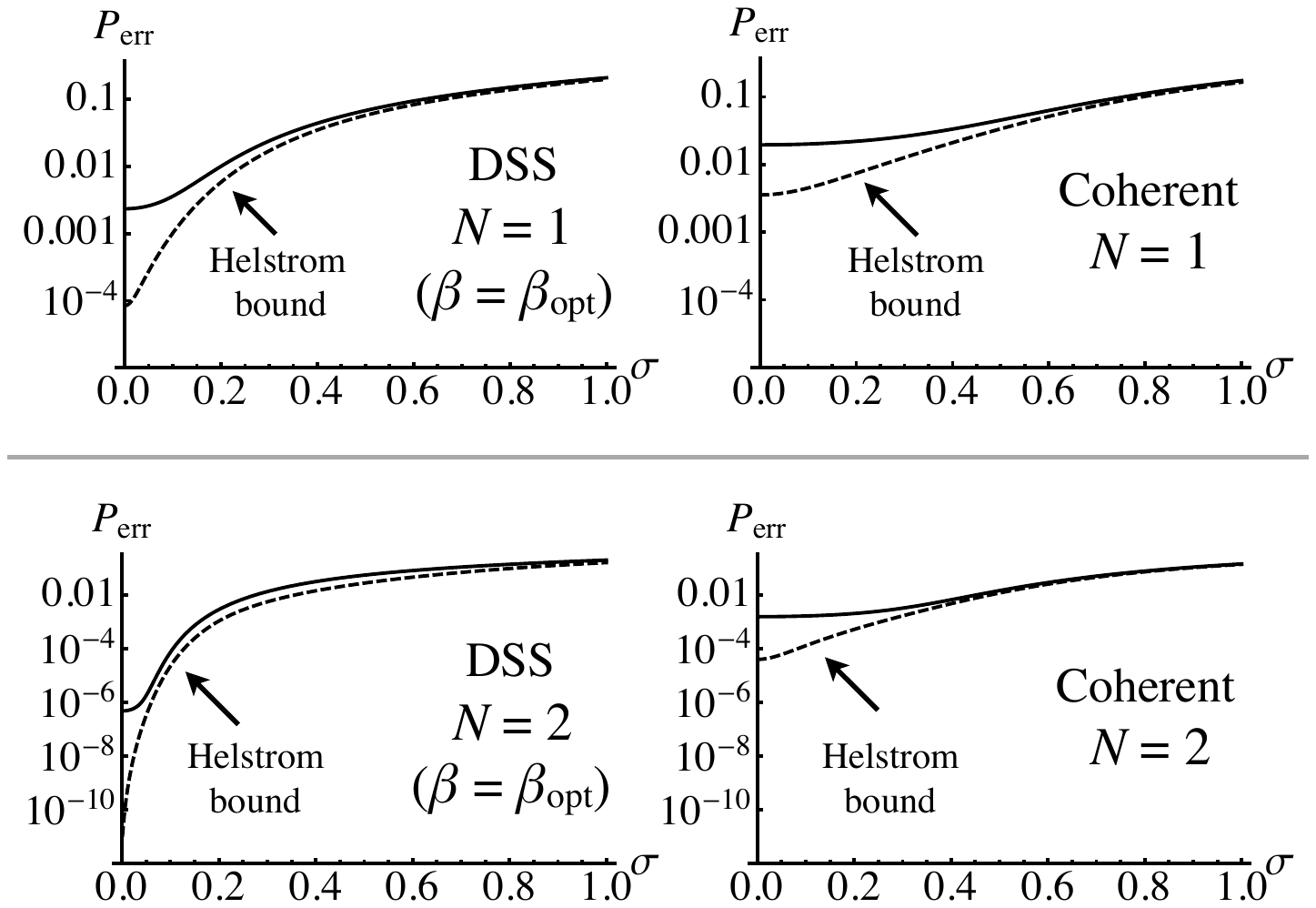}
\caption{Comparison between the error probability $P_{\rm err}$ (solid lines) and the Helstrom bound (dashed lines)
as functions of the noise parameter $\sigma$ for the DSS and the coherent state and for two different values of the
energy: (top) $N=1$ and (bottom) $N=2$. It is worth noting that, in the presence of small values on $\sigma$,
DSSs can beat the Helstrom bound of the corresponding coherent-state channel. For the DSS
we used the optimal squeezing fraction $\beta_{\rm opt}(N)$ given in Eq.~(\ref{beta:th:opt}).}
\label{fig:comparison:HBs}
\end{figure}

In Fig.~\ref{fig:comparison:HBs} we compare the Helstrom bound for the DSS with the error probability obtained by
using the homodyne strategy (plots on the left side of Fig.~\ref{fig:comparison:HBs}).
As in the case of the coherent states analysed in Ref.~\cite{olivares2}, also when the binary
information is encoded in DSSs the homodyne detection approaches the optimal measurement in the presence of phase
diffusion. It is however worth noting that DSSs allow to obtain a lower error probability and also to beat the Helstrom bound
of the coherent-state case for small values of the noise parameter $\sigma$ (see the right side of Fig.~\ref{fig:comparison:HBs}).
\subsection{Effect of purity on the discrimination}\label{sec:purity}
\begin{figure}[tb!]
\includegraphics[width=0.51\columnwidth]{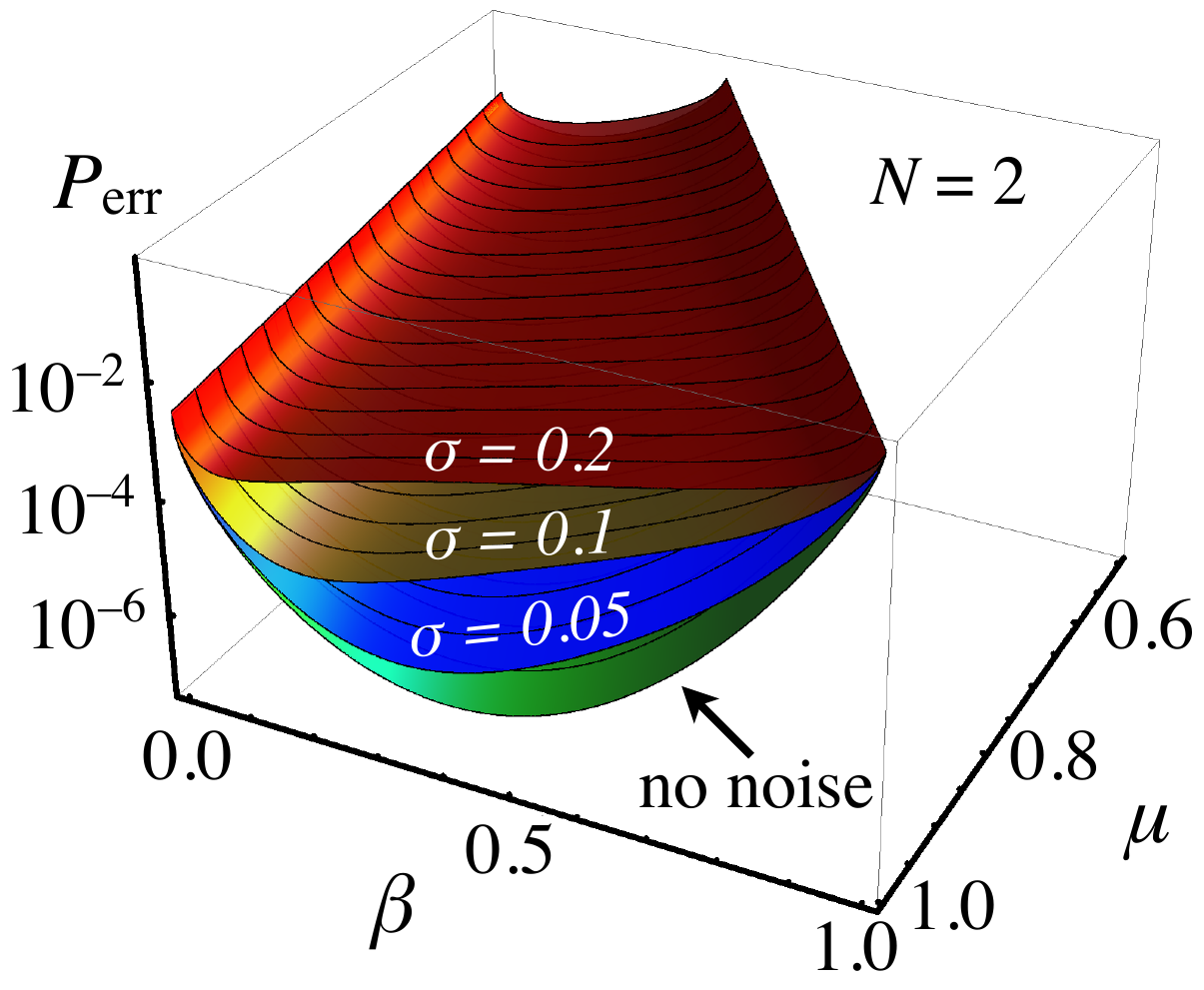}
\includegraphics[width=0.45\columnwidth]{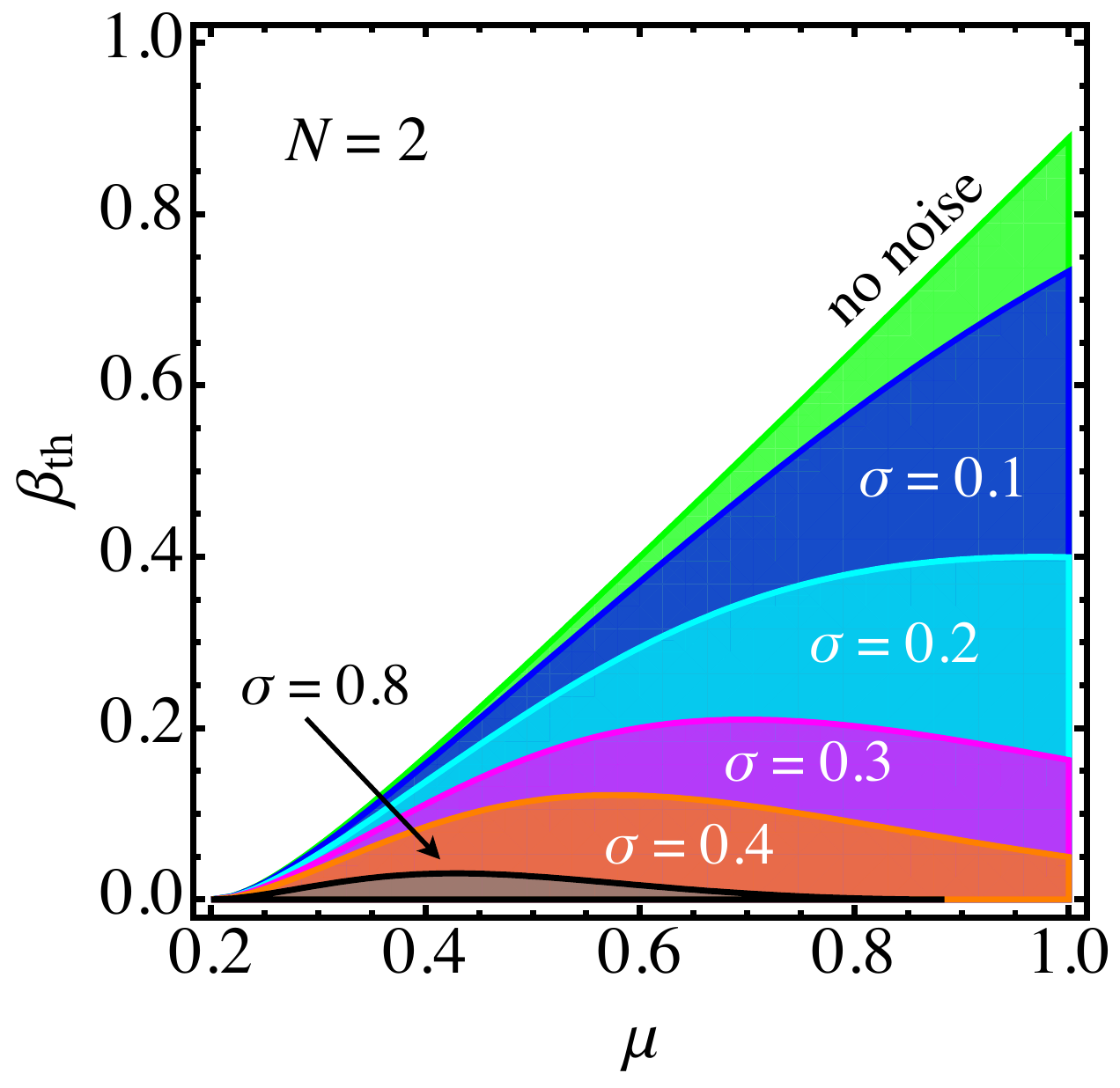}
\caption{(Left) Error probability $P_{\rm err}$ of the homodyne receiver
as a function of $\beta$ 
and the purity $\mu$ of the seed state for different values 
of the noise parameter $\sigma$. We set $N=2$. (Right) 
Threshold value $\beta_{\rm th}(\mu)$ as a function of 
the purity $\mu$ of the states for different values of the 
noise parameter $\sigma$.
The shaded regions refer to 
the pairs of parameters $(\mu,\beta)$ for which DDSs 
outperform coherent states. Also here we set $N=2$.
Note that $(1+2 N)^{-1} \le \mu \le 1$.}
\label{f5}
\end{figure}

The analysis we pursued in the previous sections focused on a 
{\em pure} seed state, namely the squeezed vacuum $| \psi_0 \rangle 
=  {S}(r) | 0 \rangle$. In order to take into account possible 
imperfections in the generation of
squeezed states and losses in their propagation, we extend our 
analysis to include the effect of a non-unit purity
$\mu= {\rm Tr}[ {\rho}_0^2]$ of the seed state which is 
now described by the density operator $ {\rho}_0$.
Without loss of generality, we can assume that in this case 
the seed is given by the {\em squeezed thermal
state} \cite{olivares:rev}:
\begin{equation}\label{degraded}
{\rho}_0 =  {S}(r)  {\nu}(\mu)  {S}^{\dag}(r)
\end{equation}
where $ {\nu}(\mu)$ is a thermal state with average 
number of photons $N_{\rm th} = (1-\mu)/(2\mu)$.
It is worth noting \cite{cialdi:sq:16,dauria:09} that this is also equivalent to the scenario in
which a \emph{pure} squeezed vacuum state $| \psi_0 \rangle 
=  {S}(\tilde{r}) | 0 \rangle$, corresponding to
an initial squeezing of $10\, \log_{10} (e^{2|\tilde{r}|})$~dB, is sent through a channel with
loss parameter $\eta$. In this case, the output state is
given by Eq.~(\ref{degraded}) with:
\begin{equation}
\mu = \frac{e^{2\tilde{r}}}{[\eta+(1-\eta)e^{2\tilde{r}}] [1+\eta(e^{2\tilde{r}}-1)]}\,,
\end{equation}
and
\begin{equation}
e^{2r} = e^{\tilde{r}}\,\sqrt{\frac{1+\eta(e^{2\tilde{r}}-1)}{\eta + (1-\eta)e^{2\tilde{r}}}}\,.
\end{equation}
In this case we can still follow the discrimination strategy mentioned above, but now the total energy of the
displaced input state $ {D}(\pm\alpha)  {\rho}_0  {D}^\dag(\pm\alpha)$ reads:
\begin{equation}
N = |\alpha|^2 + N \beta + \frac{1-\mu}{2\mu} \, (1 + 2 N \beta),
\end{equation}
where the squeezing fraction $\beta$ is the same as in Eq.~(\ref{sq:frac}). Note that $(1+2 N)^{-1} \le \mu \le 1$.
In the left panel of Fig.~\ref{f5}
we plot the corresponding error probability as a function of 
$\beta$ and the purity $\mu$
for $N=2$ (analogous results can be found for other energies) 
for different values of the noise
parameter $\sigma$ introduced in the previous section.
As expected, the presence of a mixed seed state ($\mu < 1$) 
increases the error probability. Nevertheless,
we can find again a threshold on the squeezing parameter below which this nonclassical resource enhances
the discrimination with respect to the coherent state 
(for the same fixed energy). The threshold now depends
also on the purity of the state as shown in the right panel of Fig.~\ref{f5}.

\par
\begin{figure}[tb!]
\includegraphics[width=0.45\columnwidth]{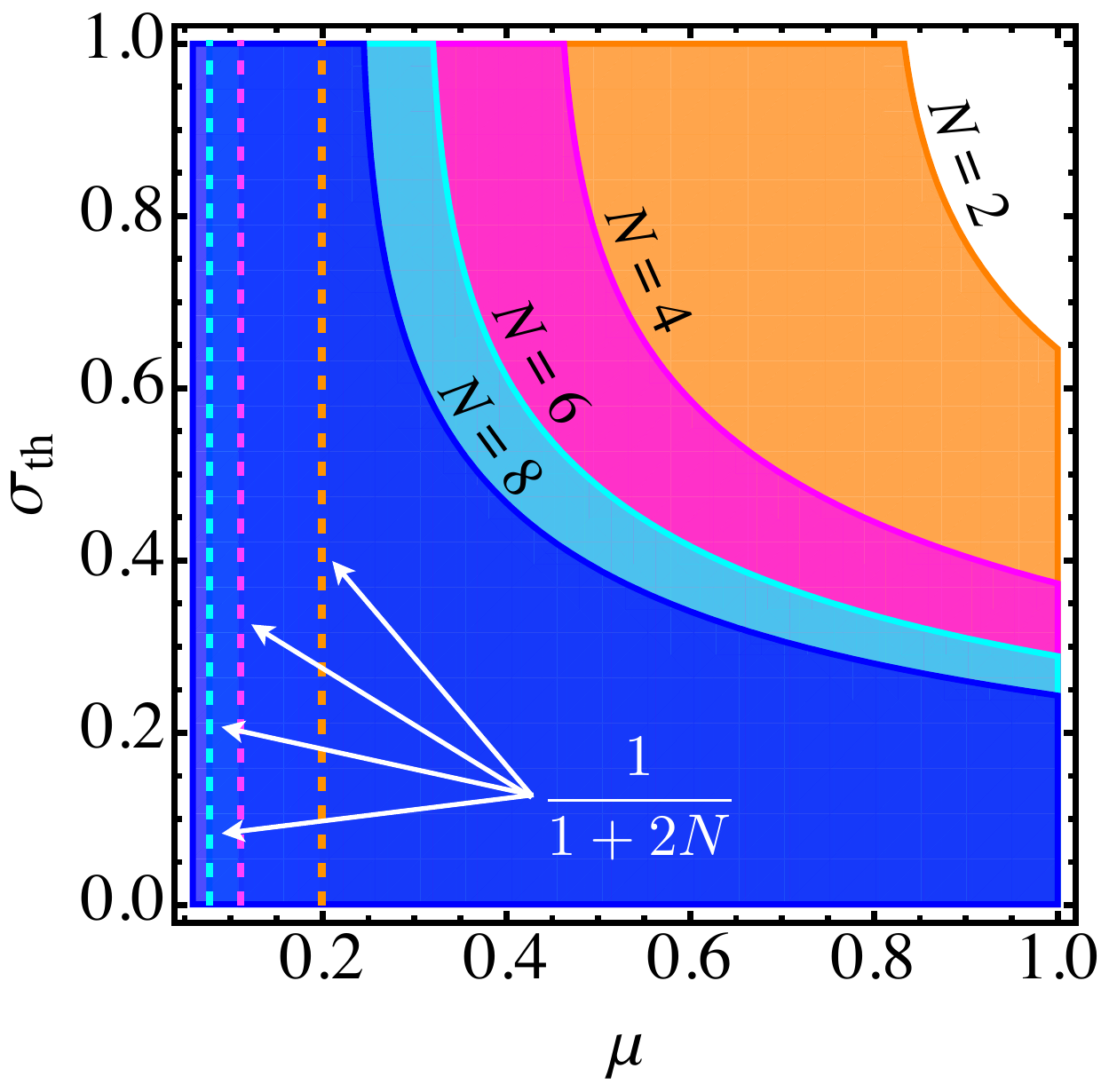}
\includegraphics[width=0.5\columnwidth]{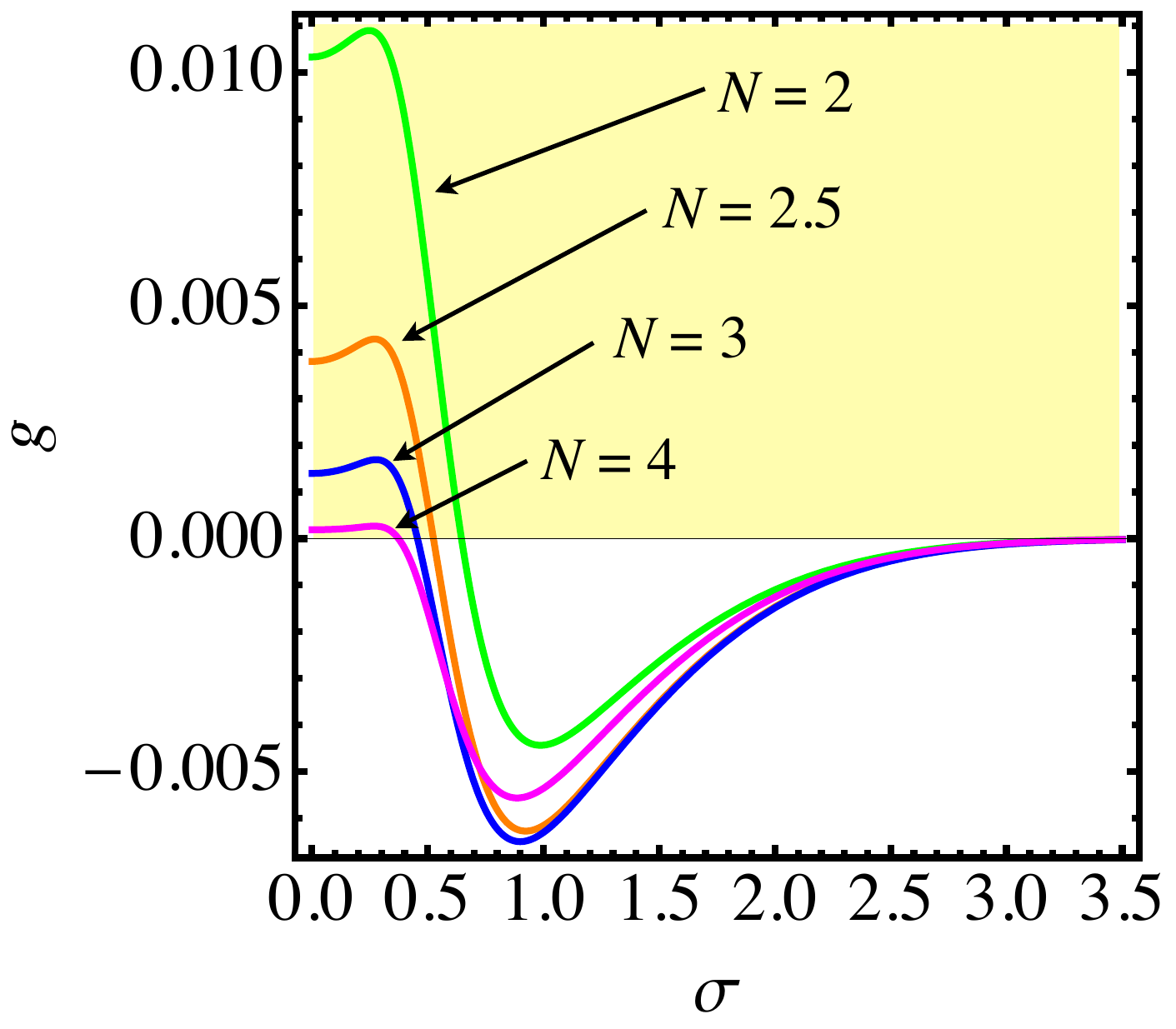}
\caption{
(Left) Plot of the threshold $\sigma_{\rm th}$ of the noise
parameter as a function of the purity $\mu$ and different values of
$N$. The vertical dashed lines refer to the lower limit of $\mu$
(see the text for details).
For $\sigma < \sigma_{\rm th}$ squeezing is useful
(of course below the corresponding $\beta_{\rm th}$).
(Right) Plot of the function $g(N;\sigma)$ in Eq.~(\ref{integrale})
as a function of $\sigma$ for different values of $N$: when its value is positive
(shaded region) we have $P_{\rm err}(\mu=1,\beta,N;\sigma) < P_{\rm err,\, CS}(N;\sigma)$
in the limit $\beta \ll 1$.
It is worth noting that as $\mu$ decreases or, equivalently, the thermal
contribution gets larger, the threshold increases. Though the corresponding
value of $\beta_{\rm th}$ becomes extremely small, it does not vanish
(see, for instance, the right panel of Fig.~\ref{f5}).
See the text for details.}
\label{f6}
\end{figure}
It is interesting to note that there exists a threshold of $\sigma$
which depends on $N$ such that squeezing can help for given
values of the purity (or of the channel losses), as shown in the
left panel of Fig.~\ref{f6}. The value of $\sigma_{\rm th}$ can be obtained by
expanding the error probability
$P_{\rm err}(\mu=1,\beta,N;\sigma)$
for $\beta \ll 1$ in the case of DSSs. By considering the lowest order
of $\beta$, namely, $\sqrt{\beta}$, we have:
\begin{align}
&P_{\rm err}(\mu=1,\beta,N;\sigma) \approx
P_{\rm err,\, CS}(N;\sigma)- g(N; \sigma)\, \sqrt{\beta\, N}\,,
\end{align}
with
\begin{align}
g(N;\, & \sigma) = \nonumber \\
& \int_{-\infty}^{+\infty} d\phi\, 
\frac{\exp\left(-\frac{\phi^2}{2\sigma^2}\right)}{\sqrt{2\pi\, \sigma^2}}\,
\frac{e^{- 2 N \cos^2\phi} \cos(2\phi) \cos\phi}{\sqrt{\pi}}\,,
\label{integrale}
\end{align}
and $\sigma_{\rm th}$ is given by the maximum value of
$\sigma$ such that $g(N; \sigma) > 0$
(see the right panel of Fig.~\ref{f6}).
Note that $P_{\rm err,\, CS}(N;\sigma) \equiv P_{\rm err}(\mu=1,0,N;\sigma)$
is the error probability in the presence of coherent states ($\beta = 0$).

\section{Conclusions}\label{sec:concl}
We have investigated the role of squeezing in PSK quantum communication
and shown that it represents a resource in the presence of phase diffusion
and losses. In particular, we focused on binary encoding onto displaced 
squeezed states (DSSs) with coherent amplitudes having opposite phases. 
We have considered the ultimate limit to the error probability given 
by the the Helstrom bound, as well as a realistic scenario based on 
homodyne detection. 
\par
In the absence of noise there exists a threshold value $\beta_{\rm th}(N)$ 
on the squeezing fraction $\beta$: below this threshold the Helstrom bound 
for DSSs is lower than that obtained using coherent channels with the same 
energy, i.e. squeezing is a resource to improve discrimination. 
When phase diffusion is taken into account, the error probability 
unavoidably increases. However, we can find a threshold value 
$\beta_{\rm th}(N,\sigma)$, now depending also on the phase-noise 
parameter $\sigma$, below which squeezing still provides enhanced 
discrimination. 
\par
Moreover, we have shown that a channel with DSSs 
encoding and homodyne detection approaches the optimality in the 
presence of phase noise, i.e. it may achieve an error probability 
close to the corresponding Helstrom bound. On the one hand, this 
confirms the findings obtained with coherent encoding \cite{olivares2}. 
On the other hand, we have found that the error probability obtained 
exploiting the squeezing may fall below the ultimate
limit given by the Helstrom bound for coherent states.
\par
Our results put squeezing forward as a resource for quantum 
communication channels in realistic conditions, namely, when 
phase-noise and losses occur during the generation and the 
propagation of information carriers. They also pave the way for 
further developments in $M$-ary communication channels.

\vfill

\end{document}